# Light-ion production in the interaction of 96 MeV neutrons with carbon

U. Tippawan[a,b], S. Pomp[a], J. Blomgren[a], S. Dangtip[a,b],

C. Gustavsson[a], J. Klug[a], P. Nadel-Turonski[a,c], L. Nilsson[a,d],

M. Österlund[a], N. Olsson[a,e], O. Jonsson[d], A.V. Prokofiev[d],

P.-U. Renberg[d], V. Corcalciuc[f], Y. Watanabe[g], A. J. Koning[h].

[a]Department of Physics and Astronomy, Uppsala University, Sweden

[b]Fast Neutron Research Facility, Chiang Mai University, Thailand

[c]Catholic University of America, Washington, DC, USA

[d]The Svedberg Laboratory, Uppsala University, Sweden

[e]Swedish Defence Research Agency, Stockholm, Sweden

[f]Horia Hulubei National Institute of Physics and Nuclear Engineering, Bucharest, Romania

[g]Department of Advanced Energy Engineering Science, Kyushu University, Japan

[h]Nuclear Research and Consultancy Group, Petten, The Netherlands

**Abstract**

Double-differential cross sections for light-ion (p, d, t, $^3$He and α) production in carbon induced by 96 MeV neutrons have been measured at eight laboratory angles from 20° to 160° in steps of 20°. Experimental techniques are presented as well as procedures for data taking and data reduction. Deduced energy-differential, angle-differential and production cross sections are reported. Experimental cross sections are compared with theoretical reaction model calculations and experimental data in the literature. The measured particle data show marked discrepancies from the results of the model calculations in spectral shape and magnitude. The measured production cross sections for protons, deuterons, tritons, $^3$He, and α particles support the trends suggested by data at lower energies.

PACS numbers: 25.40.-h, 25.40.Hs, 25.40.Kv, 28.20.-v

## I. INTRODUCTION

Fast-nucleon induced reactions are useful for investigating nuclear structure,

characterizing reaction mechanisms, and imposing stringent constraints on nuclear model calculations. Although carbon is a light nucleus, it can be expected that many statistical assumptions hold for nucleon-induced reactions at several tens of MeV. This is due to the sufficiently high level density at high excitation energies, such that shell effects and other nuclear structure signatures are washed out. Light nuclei also have low Coulomb barriers, implying that the suppression of charged-particle emission is weak. Therefore, nuclear reaction models for equilibrium and pre-equilibrium decay can be tested and benchmarked. In particular, the reaction $^{12}$C(n,n'3α) plays a crucial role in α-particle production [1]; nevertheless, very little direct experimental information concerning this reaction channel is available in the literature [2,3].

The growing interest in applications involving high-energy neutrons (E>20 MeV) demands high-quality experimental data on neutron-induced reactions. Examples are dosimetry at commercial aircraft altitudes and in space [4], radiation treatment of cancer [5-7], single-event effects in electronics [8,9], and energy production and transmutation of nuclear waste [10,11]. For all these applications, a better understanding of neutron interactions is essential for calculations of neutron transport and radiation effects. It should be emphasized that for these applications, it is beyond reasonable efforts to provide complete data sets. Instead, the nuclear data needed for a better understanding must come to a very large extent from nuclear scattering and reaction model calculations, which all depend heavily on nuclear models. These, in turn, are benchmarked by experimental nuclear reaction cross-section data.

Data on light-ion production in light nuclei, such as carbon and oxygen [12], are of great significance in calculations of dose distributions in human tissue for radiation therapy at neutron beams, as well as for dosimetry of high energy neutrons produced by high-energy cosmic radiation interacting with nuclei (nitrogen and oxygen) in the upper atmosphere [4,13]. When studying neutron dose effects in radiation therapy and at high altitude, it is unavoidable to consider carbon and oxygen due to them being dominant elements (18% and 65% by weight, respectively) in average human tissue.

In this paper, we present experimental double-differential cross sections (inclusive yields) for protons, deuterons, tritons, $^3$He and α particles produced by 96 MeV neutrons incident on carbon. Measurements have been performed at the cyclotron of The Svedberg Laboratory (TSL), Uppsala, using the dedicated MEDLEY experimental set-up [14]. Spectra have been measured at 8 laboratory angles, ranging from 20° to 160° in 20° steps. Extrapolation procedures are used to obtain coverage of the full angular distribution. Consequently, energy-differential and production cross sections are deduced, the latter by integrating over energy and angle. The experimental data are compared with the results of calculations using nuclear reaction codes and existing experimental data.

The present data have been acquired in a series of experiments. A sub-group of the initial experimental collaboration has previously analyzed and published data from a single experiment [15,16]. In this publication, all data including additional corroborating experiments are presented, as well as analysis routines which are significantly different from those used in Refs. [15,16]. The experimental methods are briefly discussed in Sec. II. The data reduction and correction procedures are described in Secs. III and IV, respectively. The theoretical framework is presented in Sec. V. In Sec. VI, experimental results are reported and compared with existing data. Conclusions and an outlook are given in Sec. VII.

## II. EXPERIMENTAL METHODS

The experimental setup has been described in detail previously [17,18], and therefore only a brief summary is given here. The neutron beam facility at TSL uses the $^7$Li(p,n)$^7$Be reaction (Q = -1.64 MeV) to produce a quasi-monoenergetic neutron beam [19]. The Li target used in the present experiment had a diameter of 26 mm and a thickness of 8 mm (427 mg/cm$^2$) and was bombarded with a proton beam of a few μA from the Gustaf Werner cyclotron. As a result, the neutron spectrum consisted of a peak at 95.6±0.5 MeV with an energy spread of 1.6 MeV full width at half maximum (FWHM) and a low-energy tail which was suppressed by time-of-flight techniques (see Fig. 1. in Ref. [12]). With a beam intensity of about 5 μA, the neutron flux at the reaction-target location is about $5 \cdot 10^4$ neutrons/(s·cm$^2$). The collimated neutron beam has a diameter of 80 mm at the location of the target, where it is monitored by a thin-film breakdown counter (TFBC) [20]. Relative monitoring was obtained by charge integration of the proton beam in a Faraday cup located in the proton beam dump. The two beam monitor readings were in agreement during the measurements.

The charged particles are detected by the MEDLEY setup [14]. It consists of eight three-element telescopes mounted inside a 24 cm high cylindrical evacuated chamber with 90 cm diameter. Eight telescopes are placed at 20° intervals, covering scattering angles from 20° to 160° simultaneously. The telescopes are mounted in two sets, one on each side of the beam, covering the forward and backward hemispheres, respectively. All the telescopes are mounted onto a turnable plate at the bottom of the chamber. By rotating the plate, the forward and backward sets of telescopes can be interchanged, thus permitting measurement of the differential cross section by two different telescope sets for the same laboratory angle. This beneficial feature guarantees data for each angle also in the case of detector malfunctioning, and it allows some telescopes to be calibrated by reasonably sharp peaks, corresponding to resolved states in (n,p) and (n,d) reaction spectra at forward angles.

Each telescope consists of two fully depleted $\Delta E$ silicon surface barrier detectors and a CsI(Tl) crystal. The thickness of the first $\Delta E$ detector ($\Delta E_1$) is either 50 or 60 μm, while the second one ($\Delta E_2$) is either 400 or 500 μm. They are both 23.9 mm in diameter (nominal). The cylindrical CsI(Tl) crystal, 50 mm long and 40 mm in diameter, serves as the $E$ detector. The back-end part of the crystal, 20 mm long, has a conical shape, tapered off to 18 mm diameter, to fit the size of a read-out diode.

To obtain a well-defined acceptance, a plastic scintillator collimator is placed in front of each telescope. A conventional collimator, thick enough to stop 100 MeV protons, can cause problems like in-scattering or particle reactions before reaching the first detector. To avoid such complications, the plastic scintillator was used as an active anti-coincidence collimator to discard the signals from particles that did not pass straight into the first detector. The plastic scintillator collimator has a 40 × 40 mm² square shape, with a 19 mm diameter hole at the center and a thickness of 1 mm. This thickness is sufficient also for the most penetrating 100 MeV protons to produce a reasonable pulse height.

Two different (cylindrical) disks of graphite are used as the carbon targets. The target diameters were 25 mm and 22 mm, with thicknesses of 150 μm and 500 μm, respectively. Each target is suspended in a thin aluminum frame using thin wires.

The dimensions of the frame have been chosen in such a way that it does not interfere with the incident neutron beam.

For absolute cross section normalization, a 25 mm diameter and 1.0 mm thick polyethylene $(CH_2)_n$ target is used. The *np* cross section at 20° and 40° laboratory angles provide the reference cross sections [21].

Instrumental background is measured by removing the target from the neutron beam. It is dominated by protons produced by neutron beam interaction with the beam tube and reaction chamber material, especially at the entrance and exit of the reaction chamber and in the telescope housings. Therefore, the telescopes at 20° and 160° are most affected.

The time-of-flight (TOF) obtained from the radio frequency of the cyclotron (stop signal for the TDC) and the timing signal from each of the eight telescopes (start signal), is measured for each charged-particle event.

The data taking was performed in two successive periods; one before rotation of the turnable plate and one after. The raw data are stored event by event for on-line monitoring and subsequent off-line analysis. Typical count rates for target-in and target-out runs were 10 and 2 Hz, respectively. The dead time of the system was typically 1-2 % and it never exceeded 10 %.

### III. DATA REDUCTION PROCEDURES

The $\Delta E$–$E$ technique is used to identify light charged particles ranging from protons to α particles, as shown in Fig.1. Good separation of all particles is obtained over their entire energy range; therefore, the particle identification procedure is straightforward. The energy resolution of each individual detector varies with the particle type [17,18]. Particles are identified by the closest-lying energy loss curve (see Fig. 1) with a maximum distance of 3σ from the tabulated values, where σ is the standard deviation of the energy resolution of each particle type.

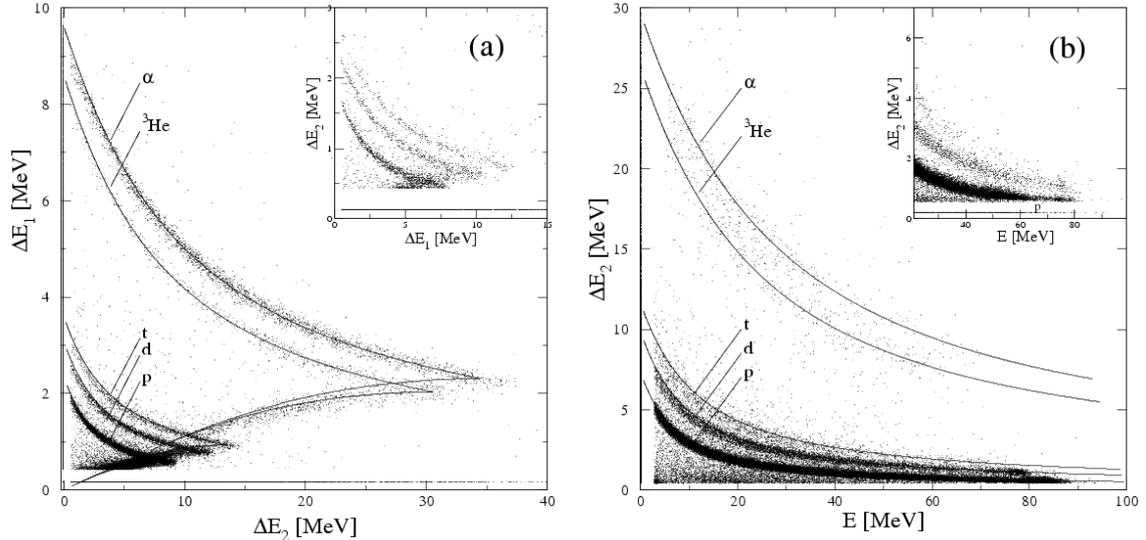

FIG. 1. Particle identification spectra at 20° for the $\Delta E_1 - \Delta E_2$ (a) and $\Delta E_2 - E$ (b) detector combinations. The solid lines represent the tabulated energy loss values in silicon [22]. The inserts in (a) and (b) illustrate the problem of the pulse-height discriminator discussed in Sec. IV.1.

Energy calibration of all detectors is obtained from the data itself [17,18]. Events in the $\Delta E$–$E$ bands are fitted with respect to the energy deposited in the two silicon detectors. This energy is determined from the detector thicknesses and calculations of energy loss in silicon (solid lines in Fig. 1). Supplementary calibration points are provided by the H(n,p) reaction, as well as transitions to the ground state and low-lying states in the $^{12}$C(n,p)$^{12}$B and $^{12}$C(n,d)$^{11}$B reactions. The energy of each particle type is obtained by adding the energy deposited in each element of the telescope.

Low-energy charged particles are stopped in the $\Delta E_1$ detector leading to a low-energy cutoff for particle identification of about 3 MeV for hydrogen isotopes and about 8 MeV for helium isotopes. The helium isotopes stopped in the $\Delta E_1$ detector are nevertheless analyzed and a remarkably low cutoff, about 4 MeV, can be achieved for the experimental α-particle spectra. These α-particle events could obviously not be separated from $^3$He events in the same energy region, but the yield of $^3$He is much smaller than the α-particle yield in the region just above 8 MeV, where the particle identification works properly. That the relative yield of $^3$He is small is also supported by the theoretical calculations in the evaporation peak region. In conclusion, the $^3$He yield is within the statistical uncertainties of the α-particle yield for α energies between 4 and 8 MeV.

Knowing the energy calibration and the flight distances, the TOF for each charged particle from target to detector can be calculated and subtracted from the registered total TOF. The resulting neutron TOF is used for selection of charged-particle events induced by neutrons in the main peak of the incident neutron spectrum (see Fig. 2. in Ref. [17]).

In order to obtain reliable differential cross sections, good knowledge of the relative solid angle is required. The solid angle acceptance is defined by the size of the collimator hole. The collimator signal from the PMT is amplified and then fed directly to a charge-sensitive ADC (QDC) for registration. A low-energy cut, corresponding to an energy loss well below that of the least ionizing particles, i.e., 100 MeV protons, is applied to the QDC spectra in the off-line analysis. Signals above this cut are used to reject the corresponding events, thus ensuring that the accepted particles passed the collimator hole.

Absolute double-differential cross sections are obtained by normalizing the target-in data to the number of recoil protons emerging from the $(CH_2)_n$ target. After selection of events in the main neutron peak and proper subtraction of the target-out and $^{12}C(n,px)$ background contributions, the cross section can be determined from the recoil proton peak, using *np* scattering data [21].

Since the target-to-detector distances and the target weights are not known precisely, some mutual normalizations and cross checks had to be introduced. In addition, comparisons with our previous experiments on oxygen and silicon [12,17,18] have to be subsequently used as a reference.

With each of the three telescopes, angular distributions for *np* scattering are measured during the experiment at three laboratory angles 20°, 40° and 60°. With the advantage of rotation of the telescope set, as described in the previous chapter, six independent normalization points are determined. Since no *np* scattering peak is visible at 80°, the target-in spectra are normalized to the *np* scattering peak at 40° by assuming the same solid angle (or the target-to-detector distances). As a cross-check, these telescopes have also been normalized using the *np* scattering peak in the 60° telescope, resulting in agreement with those normalized to the 40° telescope. Fig. 2 shows good agreement between the data sets obtained before and after rotating the turnable plate, thus they are merged to improve the statistics.

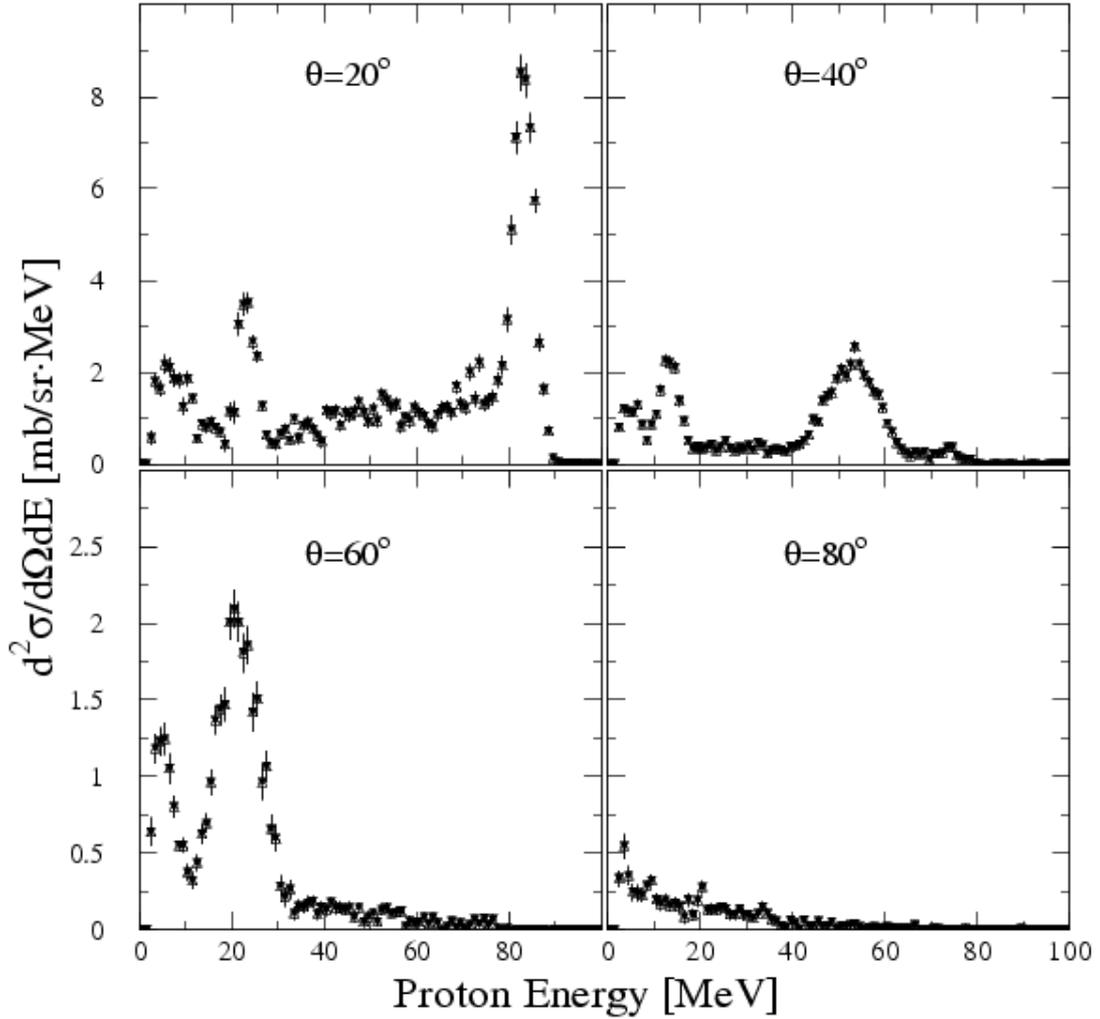

FIG. 2. Proton spectra from the CH$_2$ target at laboratory angles of 20°, 40°, 60° and 80°, resulting in six independent normalization points. The open triangles and filled upside-down triangles represent the data set before and after rotating the turnable plate, respectively.

Corroboration of the number of carbon nuclei in the graphite target and the number of hydrogen nuclei in the (CH$_2$)$_n$ target can be obtained from the ratio of the deuteron spectra between both targets on the condition that the chemical composition of the polyethylene target is known and the H(n,d) cross section at this energy is negligible. This mutual-normalization method has been applied to both thin and thick graphite targets.

As a further check, double-differential cross sections deduced from thick carbon and (CH$_2$)$_n$ targets are compared with the ones from the (CH$_2$)$_n$ target used in a similar experiment on oxygen and silicon [12,17]. Comparisons of double-differential cross sections at laboratory angles of 20°, 40°, 60° and 80° for protons and deuterons are shown in Figs. 3 and 4, respectively. The results are in agreement within their statistical uncertainties.

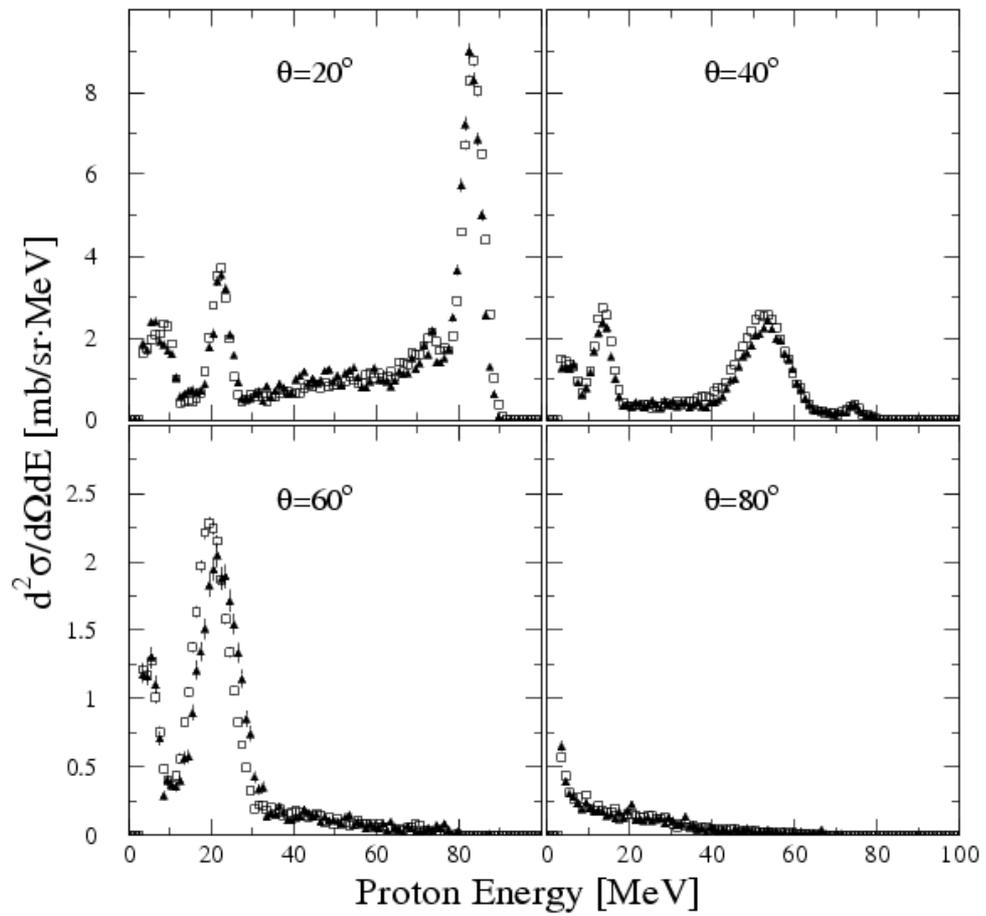

FIG. 3. Double-differential cross sections at laboratory angles of 20°, 40°, 60° and 80° for protons, deduced from the CH$_2$ targets in the present experiment (filled triangles) as well as the similar experiment (open squares) on oxygen and silicon [12,17].

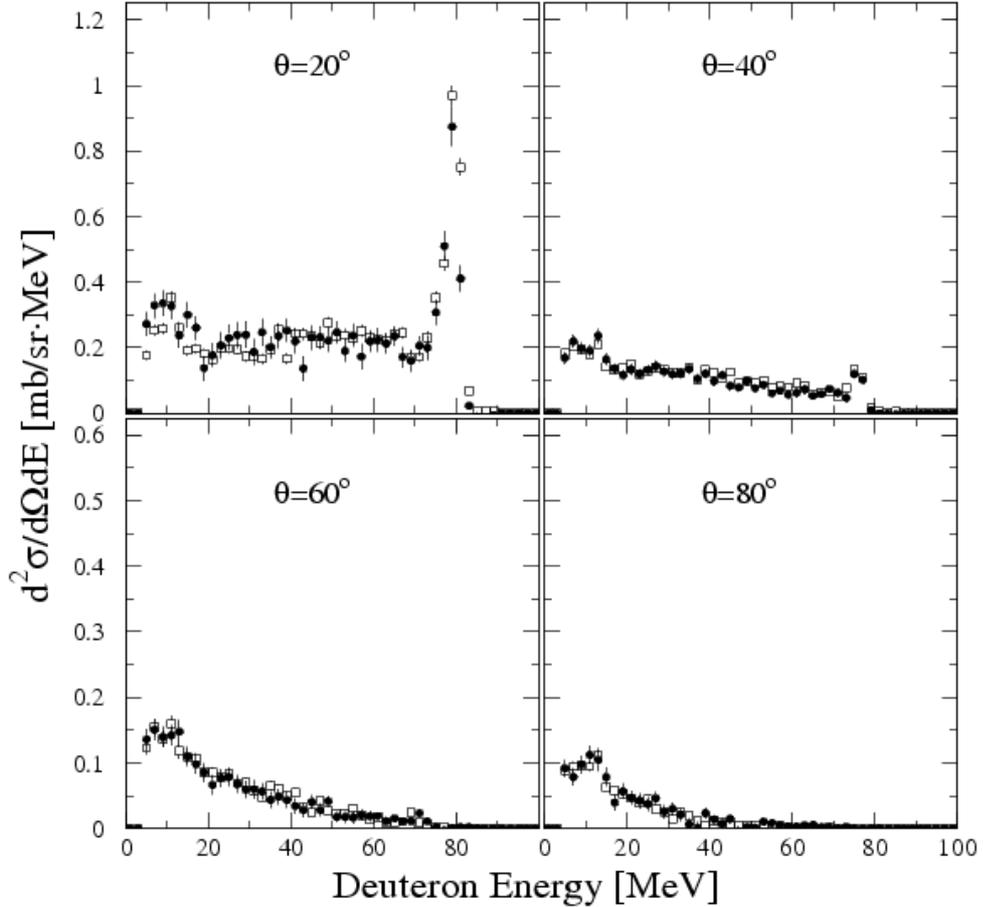

FIG. 4. Double-differential cross sections at laboratory angles of 20°, 40°, 60° and 80° for deuterons, deduced from targets of different composition. The filled circles represent the thick carbon target used in the present work, while the open squares apply to the $CH_2$ target used in the similar experiment on oxygen and silicon [12,17].

## IV. CORRECTIONS

### A. Pulse-height discriminator correction

As the ADC pulse-height discriminator in some cases was set too high, signals with amplitudes lower than the discriminator threshold have been registered as zero in pulse height. This problem causes gaps along the x- and y-axis in the $\Delta E_1$–$\Delta E_2$ and $\Delta E_2$–$E$ scatter plots (see Fig 1 (a) and (b)), whereas the cut events appear as lines at zero pulseheight.

As described in Sec. III, the low-energy cutoff for particle identification is defined by low-energy charged particles stopped in the $\Delta E_1$ detector. Particles that barely manage to punch through the $\Delta E_1$ detector, and thereafter stop in the $\Delta E_2$ detector produce signal amplitudes lower than the ADC discriminator which can be seen as a band along the y-axis of Fig. 1 (a). Thus these events cannot be distinguished from the ones that are really stopped in the $\Delta E_1$ detector. In this experiment, the gaps in each particle band lead to new energy cutoffs, higher than the usual ones [12,17] by about 0.5 MeV. For the experimental α-particle spectra including the events stopped in the $\Delta E_1$ detector, it was simply solved by making the energy bin wider in the problematic region.

In the case of high energy protons for which the deposited energy in the $\Delta E_2$ detector is rather small, the proton band was cut at high energies as seen in the inset of Fig. 1 (b). This problem can be solved easily by adding the events in the line in the region where the proton band is cut and restoring the energy loss in the second silicon detector, calculated from the tabulated energy loss referring to the energy deposited in the CsI detector. In addition, TOF and collimator QDC information and background subtraction help to eliminate unwanted events. The resulting spectra after the correction are shown in Fig. 3, which are in agreement with the corresponding experimental oxygen and silicon data [12,17].

The worst case happens when particles barely manage to pass through the second silicon detectors and stop in the CsI detectors, because then the deposited energies in both detectors are rather small. Not only energy information is missing, but the particle identification is also ambiguous. Since the particle bands turn back into the same track after punching through (see the left bottom corner of Fig. 1(a)), the particle type is identified by the $\Delta E_2$–$E$ scatter plot in Fig. 1 (b) instead. Luckily, the channel-number range of the silicon detector signals (8192 channel numbers) is twice the range of the CsI ones (4096 channel numbers), while the energy range of the silicon detector is much narrower than for the CsI crystal. As a result, the unregistered events in the CsI do not overlap each other in the second silicon detector and can also be identified by manual cuts on the zero line along the y-axis in Fig. 1 (b). Moreover, the missing energy is restored from the tabulated energy loss referring to the energy deposited in the first silicon detector. The resulting spectra after the correction are shown in Figs. 3 and 4, which are in agreement with the similar experiment on oxygen and silicon [12,17]. Nonetheless, this method cannot be used in one of the telescopes because the discriminator cuts into the proton band before the back-bending point, as seen in the inset of Fig. 1 (a). Therefore, the proton spectrum of this telescope is rejected and the data from another telescope at the same laboratory angle are used.

### B. Collimator correction

Due to malfunctioning of electronic parts, the signals from some collimators could not be used to suppress events hitting them. Therefore, a simulation program has been used to correct for this effect, as described in detail in Refs. [17,18]. As a cross check, the simulated spectra when particles punch through the plastic scintillator have also been compared with the one from the properly working collimator at the same laboratory angle. The results are in agreement within a few percent for all cases.

### C. Target thickness corrections

The advantage of the thick and thin graphite targets arrangement is two-fold. The thick target provides good statistics and the thin one enables cross-checks of the corrections of energy and particle losses due to the target thickness. These effects are calculated by a computer code, TCORR [23] which is based on iterative calculations of response functions. Figure 5 illustrates $\alpha$-particle spectra before (a,b) and after (c,d) application of the code. Corrected spectra from different thicknesses of carbon targets agree within statistical errors for the whole energy range. Furthermore, corrected spectra for targets with different carbon compositions, like graphite and polyethylene, are in good agreement (see Fig. 6).

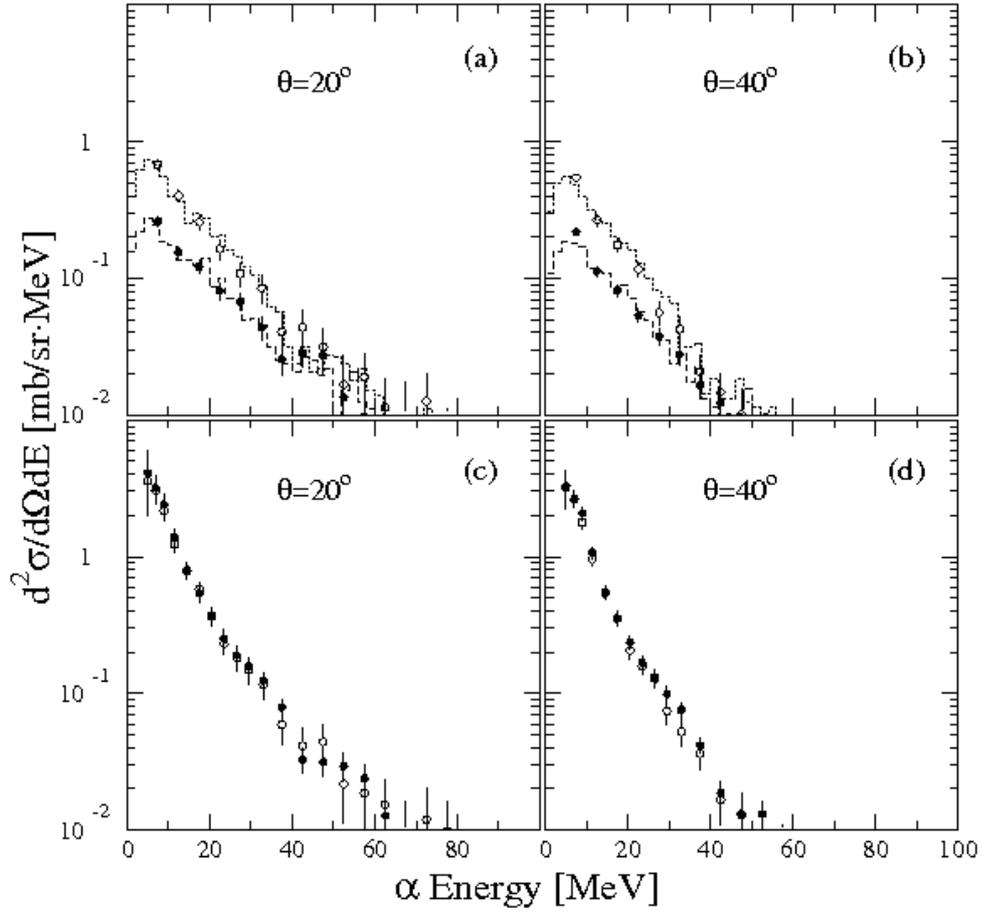

FIG. 5. α-particle spectra at laboratory angles of 20° and 40° before (a,b) and after (c,d) application of the thick target correction, TCORR [23]. The open and filled circles represent the thin and thick carbon targets, respectively. The dotted and dashed histograms in (a) and (b) show α spectra simulated for both target cases from the corrected data (see text).

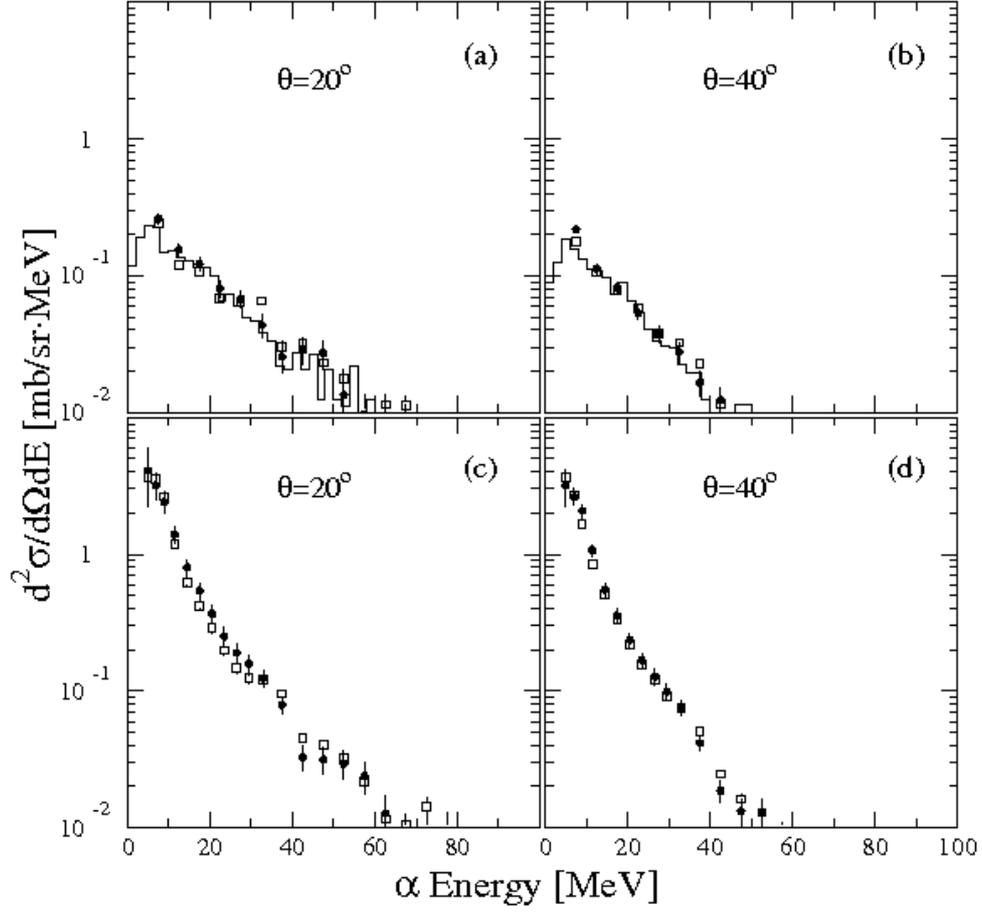

FIG. 6. Similar to Fig. 5, but the open squares represent $CH_2$ targets from the corresponding experimental oxygen and silicon data [12,17]. The solid histograms in (a) and (b) show α spectra simulated for the $CH_2$ target from the corrected data (see text).

Verification of the results from the correction method was conducted with an independent Monte Carlo program called TARGSIM [23], based on the GEANT code [24]. This program simulates the measured spectra using the corrected spectra and the MEDLEY geometry as input. In this case we started with corrected α spectra of the thick carbon target, the filled circles in Fig. 5 c-d, as true spectra input to the TARGSIM. The program simulated α particles emitted from three different targets; 150 μm and 500 μm graphite, as well as 1 mm polyethylene and then obtained pseudo-experimental α spectra using the same conditions as in the experiment. The dashed and dotted histograms in Fig. 5 a-b represent pseudo-experimental α spectra simulated for thin and thick carbon targets, respectively, whereas the solid histogram in Fig. 6 a-b corresponds to the $CH_2$ target. The simulation results reproduce the experimental data within the statistical errors over the whole energy region.

### D. Other corrections

Corrections for TOF shift and wrap-around problems are performed in analogy with the similar experiment on oxygen and silicon and are described in detail in the corresponding publications [12,17]. The data and method for the efficiency correction of the CsI(Tl) detectors, reported in Ref. [19] and used in Ref. [12,17] as well as in the present work, have recently [25] been corroborated by Monte Carlo calculations.

## V. THEORETICAL MODELS

The experimental data have been compared with nuclear theory predictions, computed with the two nuclear reaction codes GNASH [26,27] and TALYS [28]. Two sets of GNASH calculations are presented, one with parameters as reported in a recent evaluation [29], and another set with modified parameters [30] as described in Sec. V.A. For practical reasons, the two calculations are designated, in this paper, by version I for the former one and version II for the latter one. GNASH I has been widely used during the last years, while GNASH II is a recent calculation developed especially for high-energy cross section evaluations, e.g., spallation reactions at energies up to several GeV. TALYS has been published [28], however, its scope covers medium-weight nuclides and upward, thus it is described in some detail below for this particular case.

### A. GNASH calculations

The GNASH II calculation is basically the same as used for the high-energy nuclear data evaluation of the JENDL/HE-2004 file [31]. The calculation procedure is described in Ref. [30].

Nucleon transmission coefficients needed for the GNASH input were calculated using the OPTMAN code [32] based on the coupled-channels (CC) method with the nuclear Hamiltonian parameters determined by the soft-rotator model (SRM) [33]. Transmission coefficients for other light ions (d, t, $^3$He and $\alpha$) were calculated by the ECIS code [34] with the following global optical parameters: Daehnick et al. [35] for deuteron, Watanabe [36] for triton, and Ingemarsson et al. [37, 38] for $^3$He and $\alpha$. The Ignatyuk level density formula [39] was employed with a default parameter set in the statistical decay calculation.

In the GNASH II calculation, some modifications were made to the preequilibrium exciton model calculation. The Kalbach normalization factor was determined by analyses of (p, xp) and (p, nx) spectra for energies up to 150 MeV. The surface effect was taken into account in preequilibrium two-nucleons emission in the same way [40] as in single-nucleon emission, such that the same hole state density was used in a consistent way in both processes. The direct pick-up components of deuteron, triton, and $^3$He calculated using a phenomenology [41] were adjusted to provide good agreement with experimental DDX data for 68 MeV proton incidence [42] and the same normalization factors were used in the present calculation. The knockout component was ignored. The component with the exciton number 3 for deuteron preequilibrium emission was ignored and replaced by the DWBA cross section for the direct pick-up transition to the ground state, which was calculated by the DWUCK4 code [43].

The double-differential production cross sections of emitted light ions should be given in the laboratory (lab) system to compare with the present measurement. The GNASH code outputs the angle-integrated emission spectra in the center of mass (c.m.) system. The c.m.-to-lab transformation using the two-body kinematics of one-particle emission [27] is approximate when applied to the whole emission spectra including multiparticle emissions, because the velocity boost used for this transformation is valid only for the first particle emission but not for the successive decays. For large velocity boost, which is the case for particle emission from light targets like carbon, the approximation is crude, particularly for particle emission with low energies. In the GNASH II calculation, therefore, the c.m.-to-lab transformation was carried out using an empirical prescription that the moving source model [44] and the Kalbach systematics [45] were applied to the evaporation and preequilibrium components, respectively.

### B. TALYS calculations

The purpose of TALYS [28] is to simulate nuclear reactions that involve neutrons, photons, protons, deuterons, tritons, $^3$He and α-particles in the 1 keV – 200 MeV energy range. Predicted quantities include integrated, single- and double-differential cross sections, for both the continuum and discrete states, residue production and fission cross sections, gamma-ray production cross sections, etc. For the present work, single- and double-differential cross sections are of interest. To predict these, a calculation scheme is invoked which consists of a direct + pre-equilibrium reaction calculation followed by subsequent compound nucleus decay of all possible residual nuclides calculated by means of the Hauser-Feshbach model.

Obviously, a target nucleus as light as $^{12}$C is a particular case, and theoretically beyond the validity range of nuclear models such as the optical model, level density and pre-equilibrium model. Nevertheless, in the absence of a reliable alternative, at the high incident energy considered in this work an adequate description of the basic scattering observables is expected, at least for the incident neutron channel and the high energy inelastic scattering and charge-exchange leading to discrete states and the continuum. This situation is analoguous to the similar work reported in [12], so we only repeat here the essential ingredients and adjustable parameters needed to get a good description for $^{12}$C.

For the neutron and proton optical model potentials (OMP), the global OMP of Ref. [46] was used. Although the global neutron OMP has been validated for A > 24, for the low-energy outgoing charged particles, the invalid use of the global OMP may for such light nuclides have larger consequences. Obviously, a system of a total of 13 nucleons can hardly be called statistical, and this short-coming may be reflected in the prediction of some of the observables that concern low emission energies. For complex particles, the optical potentials were directly derived from the usual folding approach.

The default pre-equilibrium model of TALYS is the two-component exciton model [47]. A remark similar to that given above for the OMP applies: the two-component exciton model for nucleon reactions has been tested, rather successfully, against basically all available experimental nucleon spectra for A > 24 [47]. The current system A = 13 falls outside that mass range, and does not entirely qualify as a system that can be handled by fully statistical models such as the exciton model. To get the best overall description for $^{12}$C, we multiply the standard matrix element of Ref. [47] by a factor of 0.5. The partial level density parameters used are $g_\pi = Z/15$ and $g_\nu = N/15$ MeV$^{-1}$ although for $^{12}$B values of 0.8 for both $g_\pi$ and $g_\nu$ are used. Multiple pre-equilibrium processes, i.e. the emission of more than one fast particle from the non-equilibrated residual nucleus, are taken into account. The double-differential cross sections are obtained from the angle-integrated spectra using the Kalbach systematics [48]. For preequilibrium reactions involving deuterons, tritons, $^3$He and α-particles, the phenomenological model of Kalbach [49] is implemented in TALYS.

To account for the evaporation peaks in the charged-particle spectra, multiple compound emission was treated with the Hauser-Feshbach model. In this scheme, all reaction chains are followed until all emission channels are closed. For the level density, the Constant Temperature Model is used, using the global parameterization of Ref. [50].

## VI. RESULTS AND DISCUSSION

Double-differential cross sections at laboratory angles of 20°, 40°, 100° and 140° for protons, deuterons, tritons, $^3$He and α particles are shown in Figs. 7-11, respectively. All spectra for each particle type are plotted on the same cross section scale to facilitate comparison of their magnitude. The choice of the energy bin width is a compromise between the energy resolution in the experiment, the width of the inverse response functions [23] and acceptable statistics in each energy bin. The vertical bars represent statistical uncertainties only.

In order to improve the statistics, the present so-called thick-carbon data-set has been combined with the so-called CH$_2$ data-set from Ref. [12,17] (see Sec. III for details). The results are shown in Figs. 7-11. However, due to the elastic *np* scattering contribution from hydrogen nuclei in the CH$_2$ target, this is only possible for the proton spectra at backward angles. Thus, the statistical uncertainty of individual data points in the in the double-differential spectra at 20° is typically 10% for protons (as in the pure carbon data-set alone), 20% for tritons, 20% for $^3$He and 15% for α-particles. As the angular distributions are forward-peaked, these values increase with angle. The systematic uncertainty contributions are due to the thick-target correction (1-20%), beam monitoring (2-3%), particle identification (3%), CsI(Tl) intrinsic efficiency (1%) and dead time (<0.1%). The overall uncertainty in the absolute cross section is about 10%, which is due to uncertainties in the *np* scattering angle, statistics in the *np* scattering peak (5%) and the analysed uncertainties of the *np* scattering spectra which are related to the number of hydrogen and carbon nuclei (5%), and relative solid angle (5%), the contribution from the low-energy continuum of the $^7$Li(p,n) spectrum to the *np* scattering proton peak (3%), and the reference *np* cross sections (2%) [21]. The systematic uncertainties of the CH$_2$ data are quoted in Refs. [12,17].

From Figs. 7–11 it is obvious that the charged-particle emission from 96 MeV neutron irradiation of carbon is dominated by proton, deuteron and α particle channels. The spectra of the other two particle types studied in this work (tritons and $^3$He) are lower by an order of magnitude. All of the spectra have more or less pronounced peaks at low energies (below 10-15 MeV) or at least a sharp rise in the cross section towards lower energies. The angular distributions (see below) are not too far from isotropy at low energies. No low-energy peak or strong cross section rise is observed in the $^3$He spectra because of the 8 MeV low-energy cutoff discussed in Sec. III.

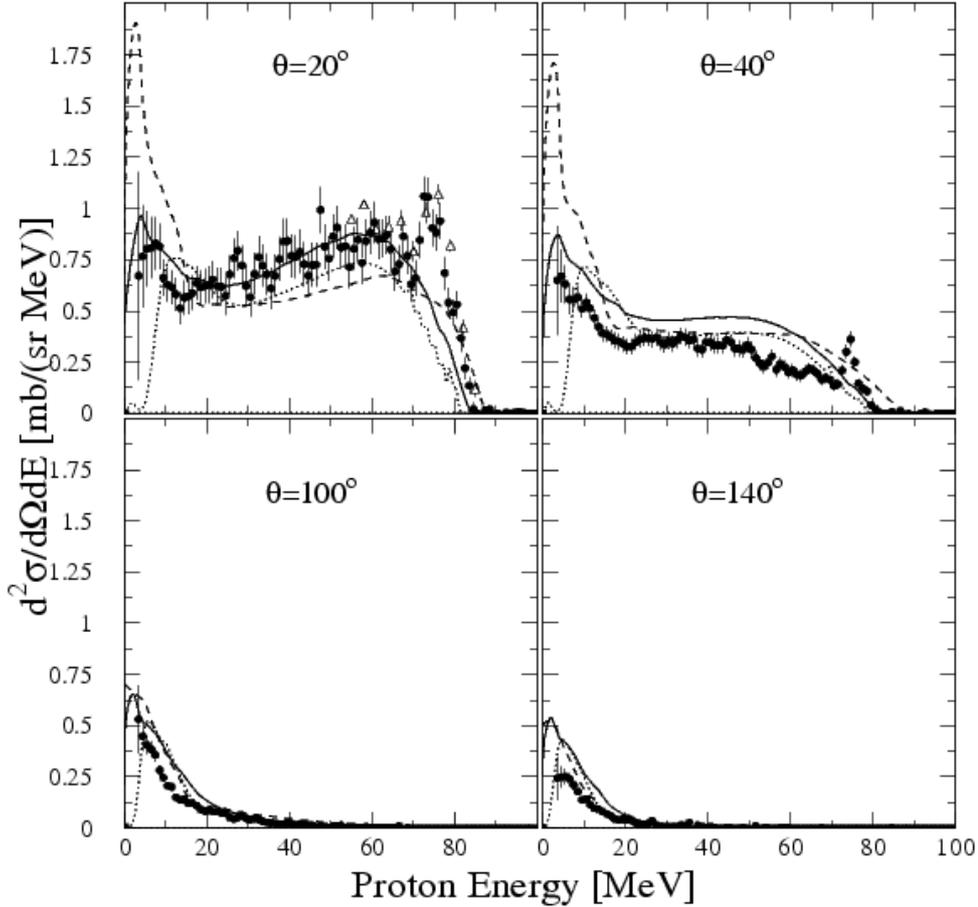

FIG. 7. Experimental double-differential cross sections (filled circles) of the C(n,px) reaction at 96 MeV at four laboratory angles. Measured data from Olsson *et al.* [51] at E$_n$ = 98 MeV (open triangles) are shown at laboratory angles of 20°. Note that the energy scale of the open triangles is shifted down by 2 MeV to facilitate comparison. Curves indicate theoretical calculations based on GNASH I [29] (dashed), GNASH II [30] (solid) and TALYS [28] (dotted).

The general trend observed is a decreasing particle-emission probability with increasing angle, over the full energy range. All particle spectra at forward angles have relatively large magnitudes at medium to high energies. The emission of high-energy particles is strongly forward-peaked and hardly visible in the backward hemisphere. It is a sign of particle emission before statistical equilibrium has been reached in the reaction process. In addition to this broad distribution of emitted particles, the deuteron spectra at forward angles show narrow peaks corresponding to transitions to the ground state and low-lying states in the final nucleus, $^{11}$B. These transitions are most likely due to pick-up of weakly bound protons in the target nucleus, $^{12}$C. Less pronounced peaks are observed in the proton and triton spectra at forward angles. The peak at about 75 MeV in the 40$^o$ proton spectrum probably corresponds to the 4.5 MeV excitation energy peak observed by Olsson et al. [51] and attributed mainly to the $2^-$ and $4^-$ states in $^{12}$B. The DWBA calculations of Ref. [51] indicate a rather flat angular distribution from 0$^o$ to 30$^o$ in the c.m. system, which explains the pronounced peak seen in the spectrum. The spectral shape and magnitude in the mid-energy region 20–60 MeV for all particle types are mostly very similar to the ones in corresponding experimental oxygen and silicon data [12,17].

**VI.1 Comparison with theoretical model calculations**

Fig. 7 shows a comparison between the double-differential (n,px) experimental spectra and the calculations based on TALYS (dotted), GNASH I (dashed) and GNASH II (solid). At 20$^o$, GNASH II gives a better description of the spectra than GNASH I and TALYS. All calculations overpredict the magnitude of the proton spectra somewhat at large angles.

Olsson *et al.* [51] have measured double-differential cross sections of the $^{12}$C(n,p)$^{12}$B reaction at $E_n$ = 98 MeV using the LISA magnet spectrometer. The data extend down to 30 MeV below the maximum proton energy and cover 0$^o$-30$^o$. Their results at 20$^o$, shown as open triangles in Fig. 7, agree very well with the present data.

At low energies, in the compound nucleus region, calculations of GNASH II agree better with the data than GNASH I. The positions of the peaks predicted by TALYS are displaced. The general decreasing trend of the evaporation peaks with decreasing mass number is seen in Ref. [52]. The decay process is less prominent because of low level density.

At the high-energy spectral part, the visible peaks at laboratory angles of 20° and 40° are most likely the result of a strong component of a direct knock-out reaction, e.g., (n,px) scattering. None of the calculations account for these peaks in the experimental data.

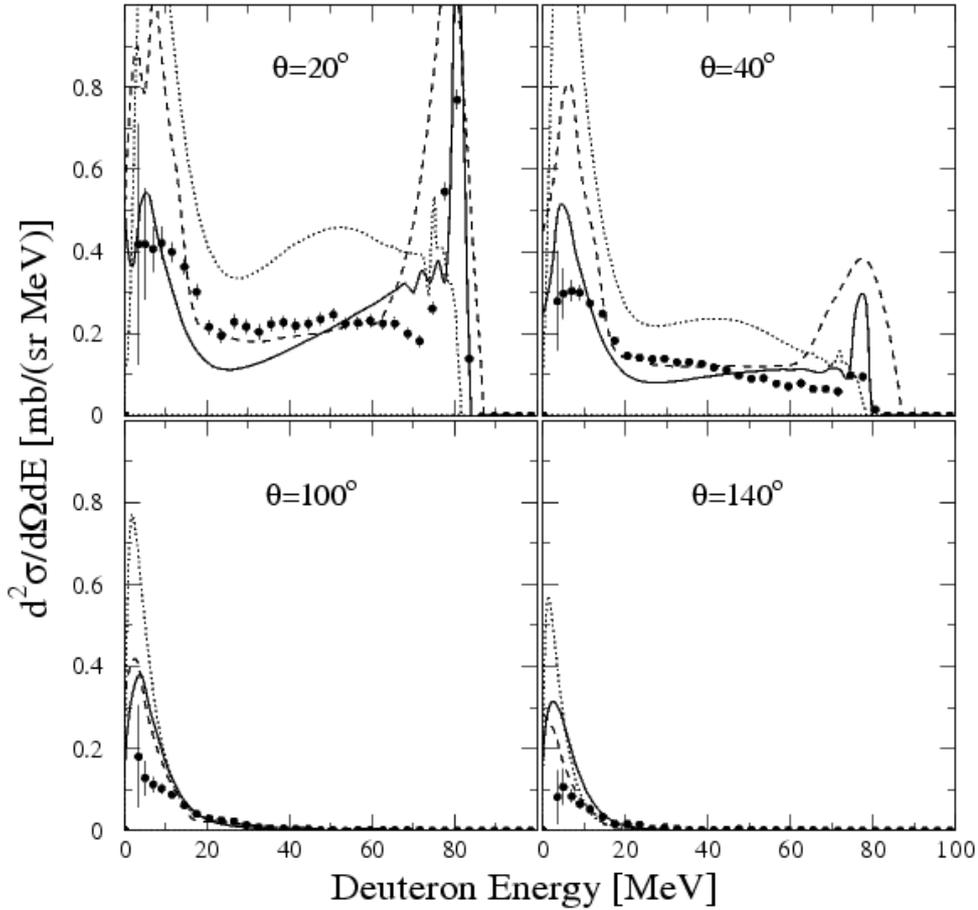

FIG. 8. Experimental double-differential cross sections (filled circles) of the C(n,dx) reaction at 96 MeV at four laboratory angles. Curves indicate theoretical calculations based on GNASH I (dashed), GNASH II (solid) and TALYS (dotted).

For the deuteron spectra (Fig. 8), none of the predictions account very well for the data. Deviations of a factor of two or more are present at all angles. At forward angles the high-energy part is strongly overestimated, indicating problems in the hole-strength treatment. There is a large difference in the spectral shapes calculated with the two versions of GNASH. This difference results from the fact that emission from the configurations with exciton number 3 is neglected in GNASH II calculations. This component is taken into account as a direct pickup component calculated with an empirical formula developed by Kalbach [45]. As seen in the proton spectra, the statistical peak is overpredicted by the GNASH I and TALYS calculations essentially at all angles, whereas GNASH II calculations seem to do a slightly better job in this case.

The overall shapes of the triton and $^3$He spectra (Figs. 9–10) are poorly described by the calculations. TALYS calculations seem to account better for the spectrum shapes. GNASH II predicts an intensity bump structure that overestimates the experimental data in the mid-energy region 30–60 MeV at forward angles. This structure is related to pre-equilibrium reactions based on the exciton model in the GNASH code. At backward angles, the yield is very small and it is difficult to make quantitative comparisons.

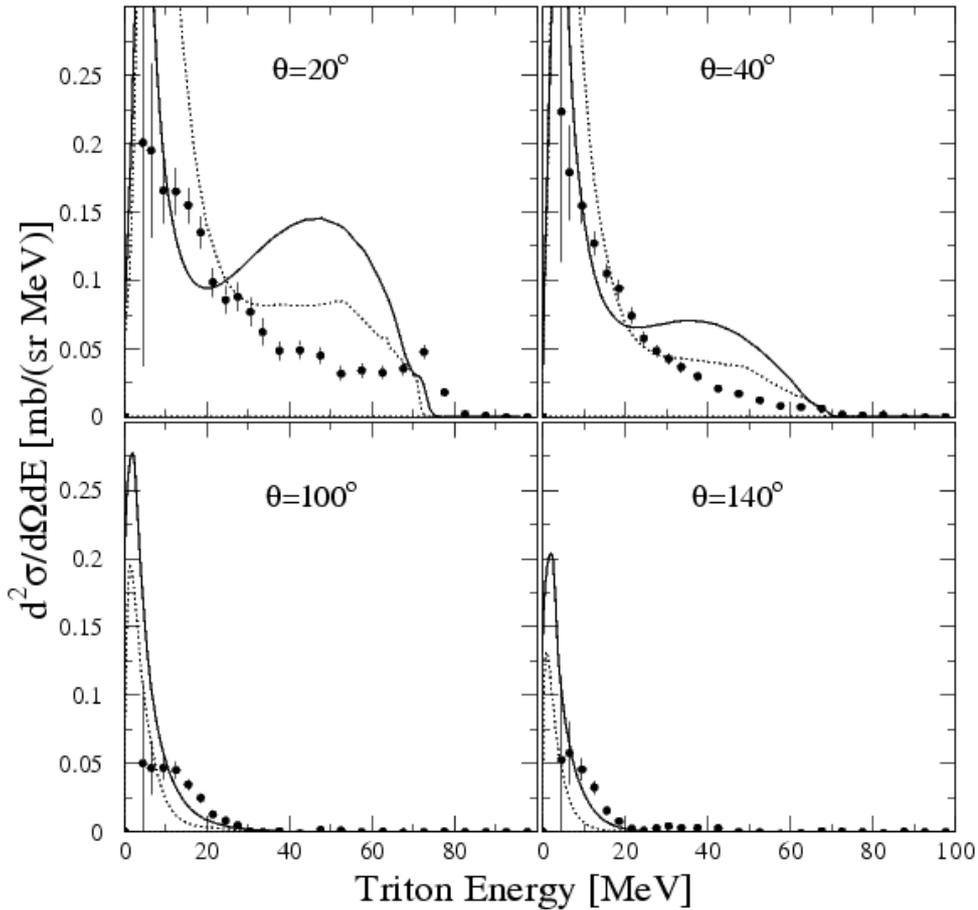

FIG. 9. Same as fig. 7, but for the C(n,tx) reaction. Note there is no calculation of GNASH I.

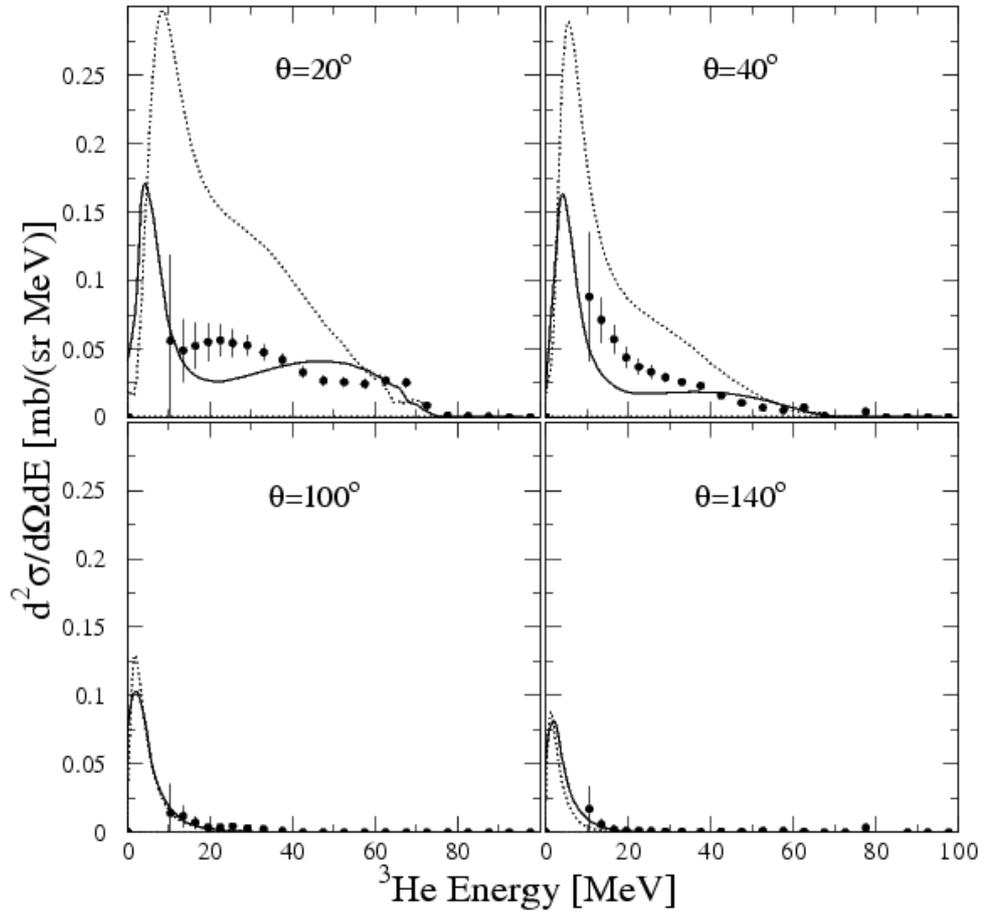

FIG. 10. Same as fig. 7, but for the C(n,$^3$Hex) reaction. Note that there is no calculation of GNASH I.

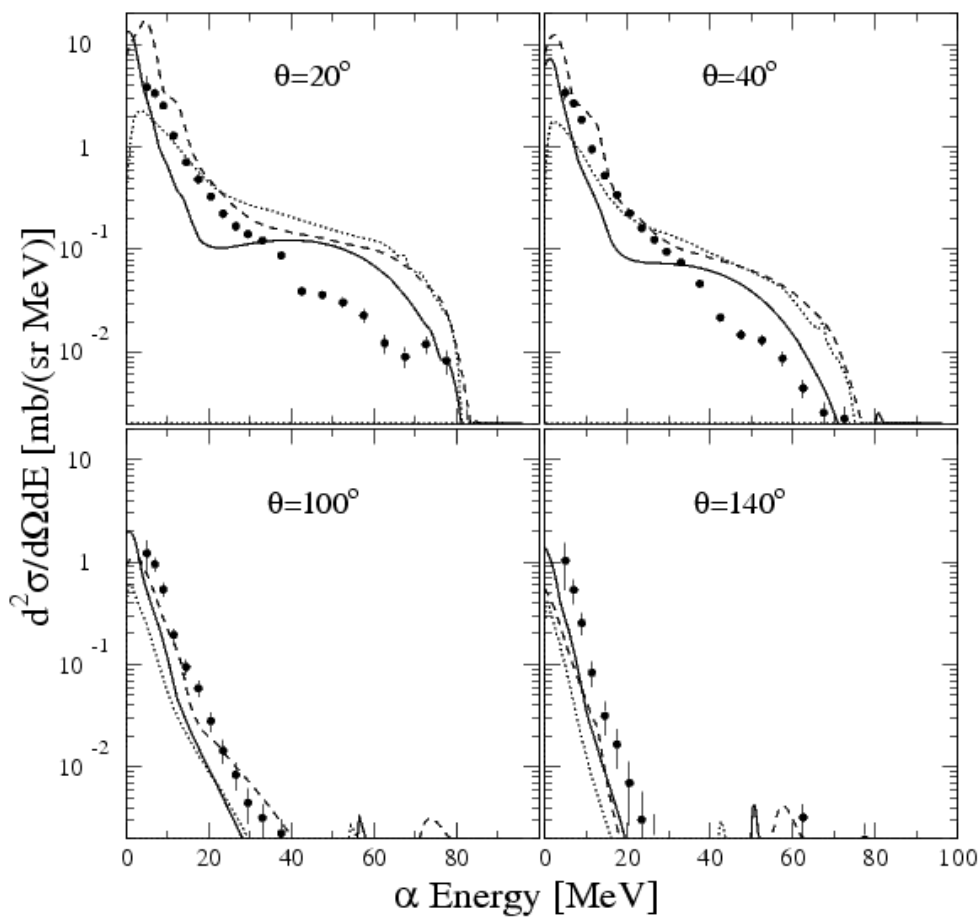

FIG. 11. Same as fig. 7, but for the C(n,αx) reaction. Note the logarithmic scale.

None of the calculations give a satisfactory description of the overall shapes of the α-particle spectra (Fig. 11). The GNASH I and TALYS codes give reasonable predictions in the energy region 15–30 MeV whereas the GNASH II code underpredicts them at forward angles. Above 30 MeV, all calculations exhibit an intensity bump structure at forward angles (but in different energy regions) whereas GNASH II and TALYS calculations underpredict at backward angles.

The ability of the models to account for the low-energy peak caused by the evaporation processes (and for α-particles also the 3α breakup of $^{12}$C) is not impressive. In general, both GNASH models tend to overpredict the cross sections whereas the TALYS model does the opposite. However, the peak maximum is close to (for $^3$He below) the low-energy cutoff, which complicates the comparison. Another complication in this context is that GNASH I and TALYS cross sections, despite being given in the laboratory system, are calculated using the kinematics of one-particle emission [26,27] for the c.m.-to-lab transformation, which is obviously an approximation while GNASH II cross sections are transformed using an empirical formula, namely the moving source model (see Sec. V).

For a detailed comparison with theoretical models, angular distributions are needed. Experimental angular distributions at low, medium and high ejectile energies are shown in Figs. 12–16 for protons, deuterons, tritons, $^3$He, and α particles, respectively. The data are compared with angular distributions calculated on the basis of the GNASH I, II and TALYS models. In general, the GNASH I and TALYS models give a steeper angular dependence than the data. The GNASH II model, despite being closer to the data, tends to give a slightly weaker angular variation, especially at low energies for helium isotopes.

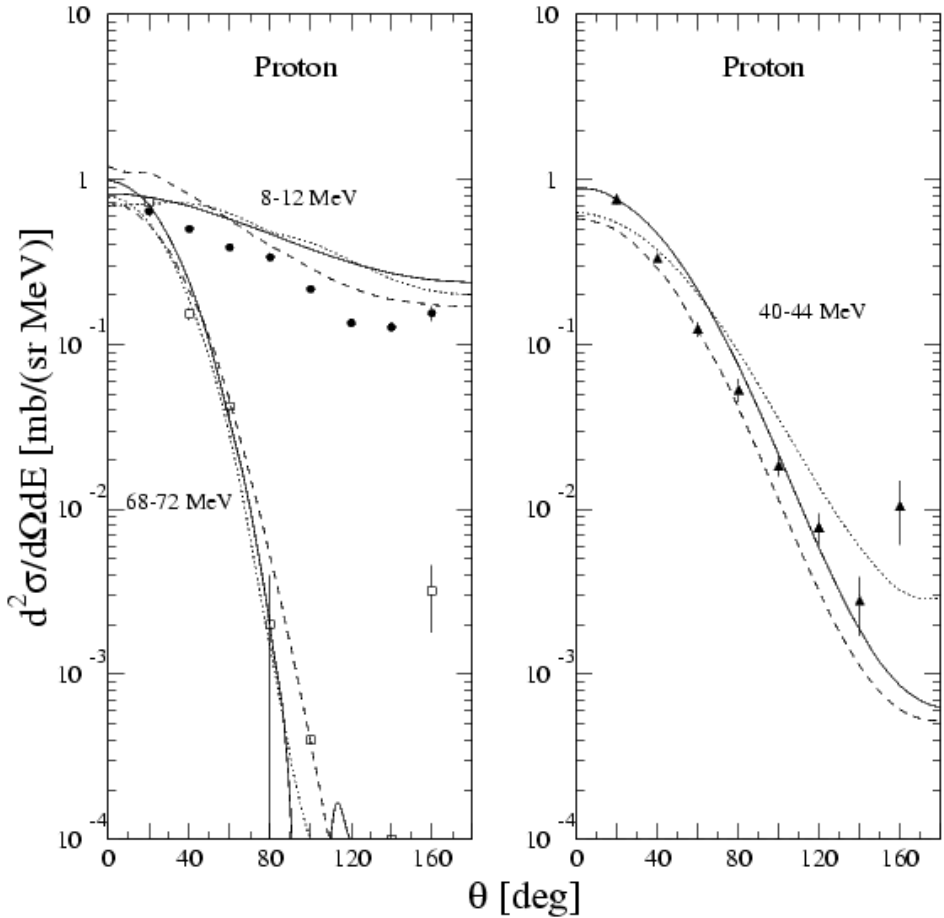

FIG. 12. Angular distributions of the C(n,px) cross section at ejectile energies of 8–12 MeV (filled circles), 40–44 MeV (filled triangles), and 68–72 MeV (open squares). Curves indicate theoretical calculations based on GNASH I (dashed), GNASH II (solid) and TALYS (dotted).

For protons and deuterons, all models give a good description of the data, except in the low-energy region, where GNASH I and TALYS calculations predict steep forward-peaked angular distribution for deuterons.

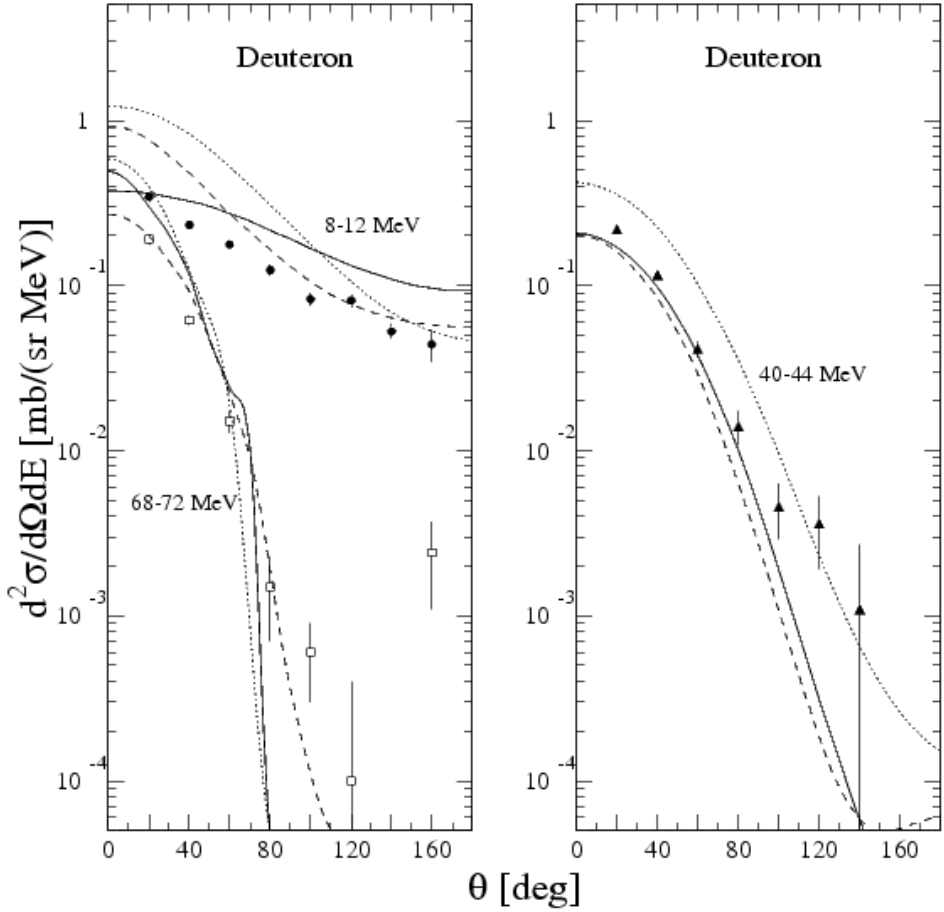

FIG. 13. Same as fig. 12, but for the C(n,dx) reaction.

The situations are similar for tritons and $^3$He, where the GNASH II describes the data well while the TALYS code overpredicts at small angles and underpredicts at large angles. The weakly forward-peaked angular distribution suggests that the tritons and $^3$He spectra at these emission energies are multistep-compound dominated, whereas those of TALYS calculations are not. None of the calculations give a satisfactory description of the magnitudes of the α-particle spectra.

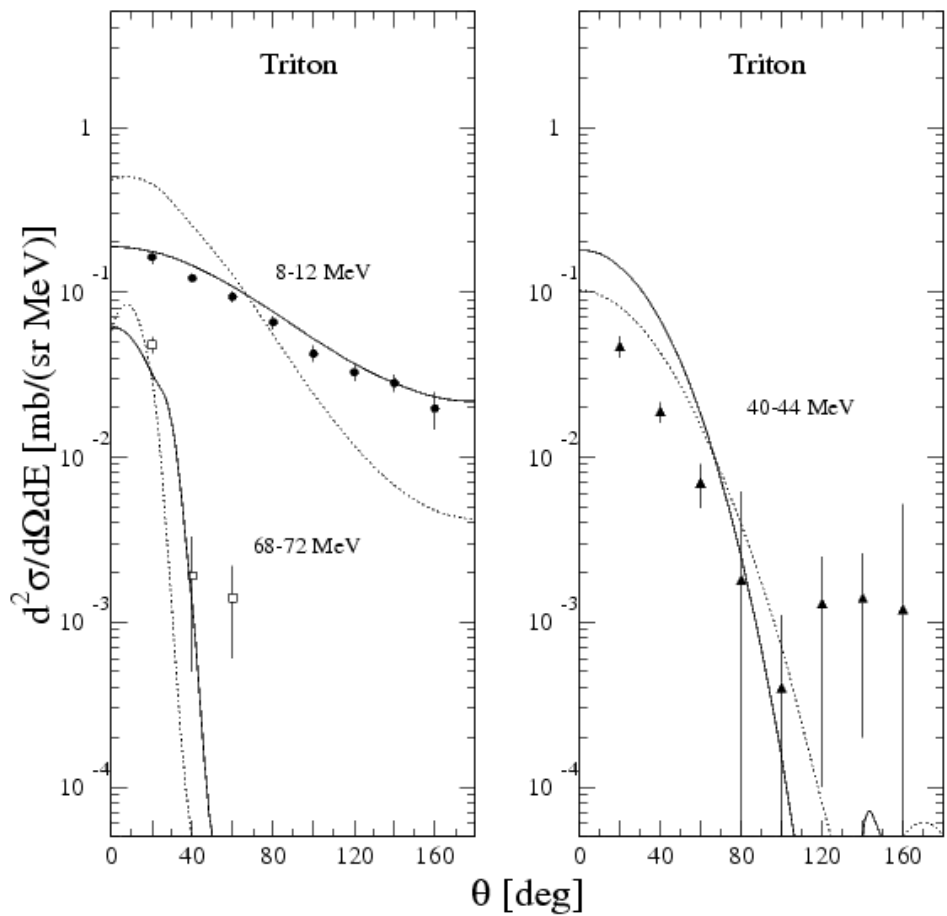

FIG. 14. Same as fig. 12, but for the C(n,tx) reaction. Note that there is no calculation of GNASH I.

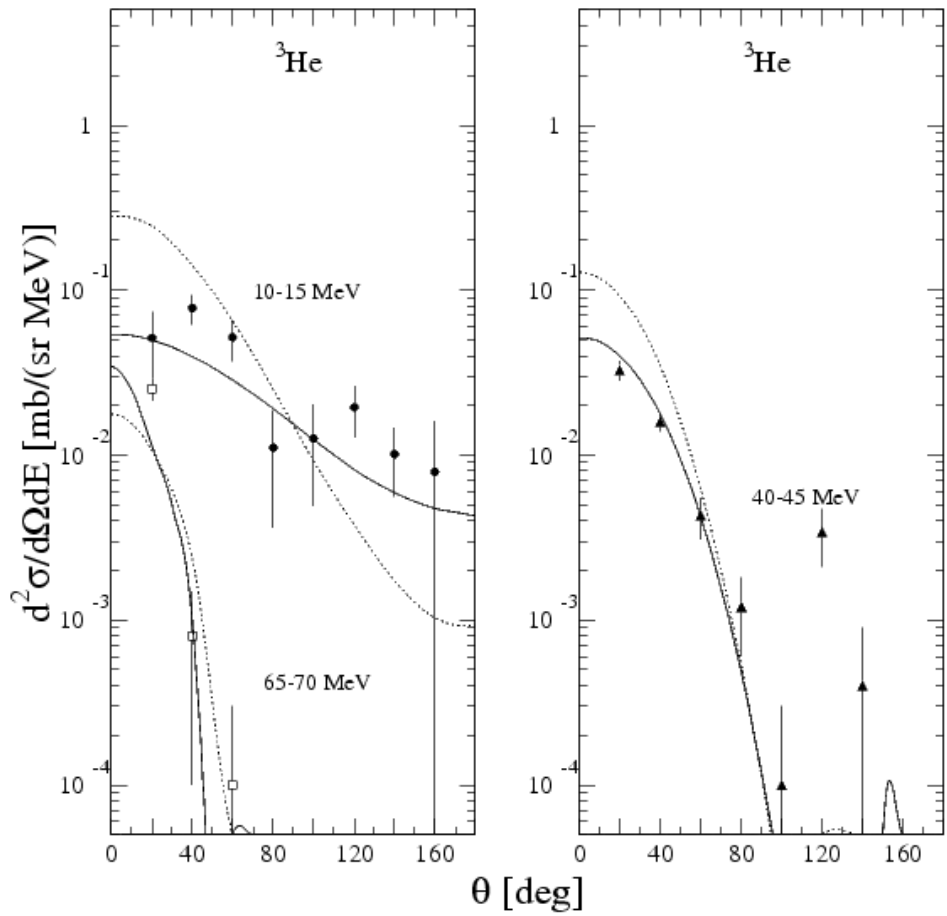

FIG. 15. Same as fig. 12, but for the C(n,³Hex) reaction. Note that there is no calculation of GNASH I.

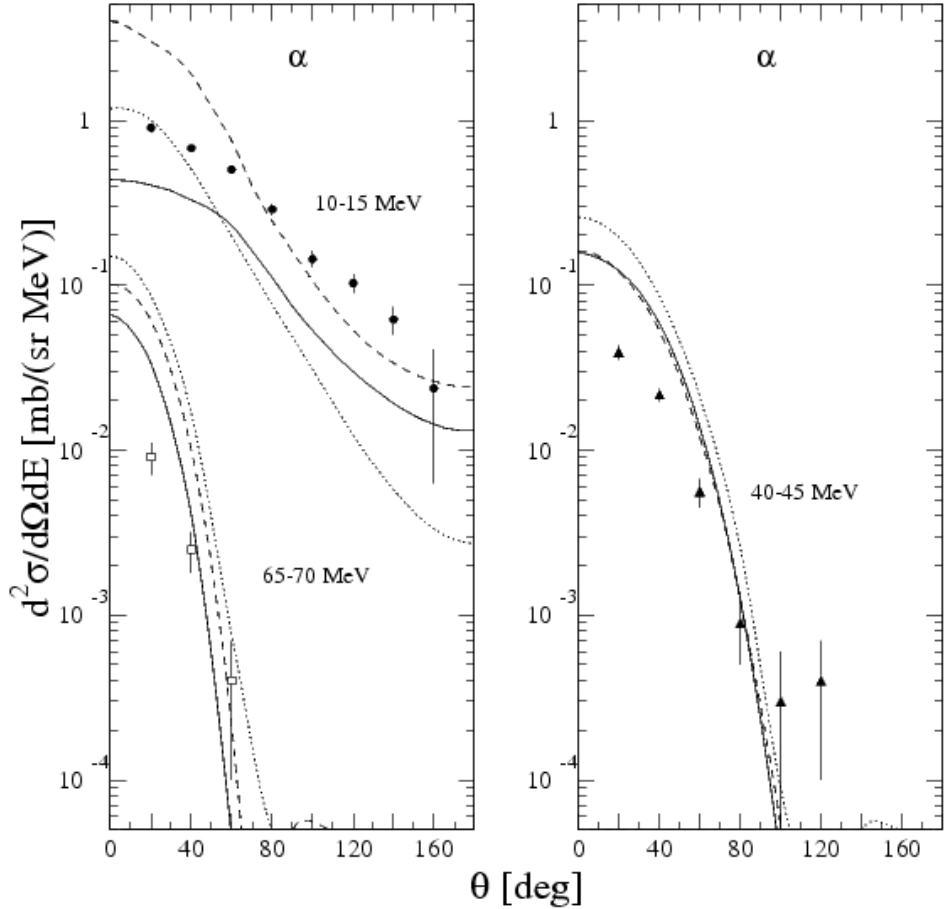

FIG. 16. Same as fig. 12, but for the C(n,αx) reaction.

**VI.2 Integrated spectra**

For each energy bin of the outgoing light charged particle spectra, the experimental angular distribution is fitted by the simple two-parameter functional form $a \exp(b \cos \theta)$ [45]. This allows extrapolation of double-differential cross sections to very forward and very backward angles. In this way, coverage of the full angular range is obtained. By integration of the angular distribution, energy-differential cross sections (dσ/dE) are obtained for each ejectile. It is applied separately to the pure carbon, the $CH_2$, and the merged data sets. These are shown in Fig. 17 together with the theoretical calculations.

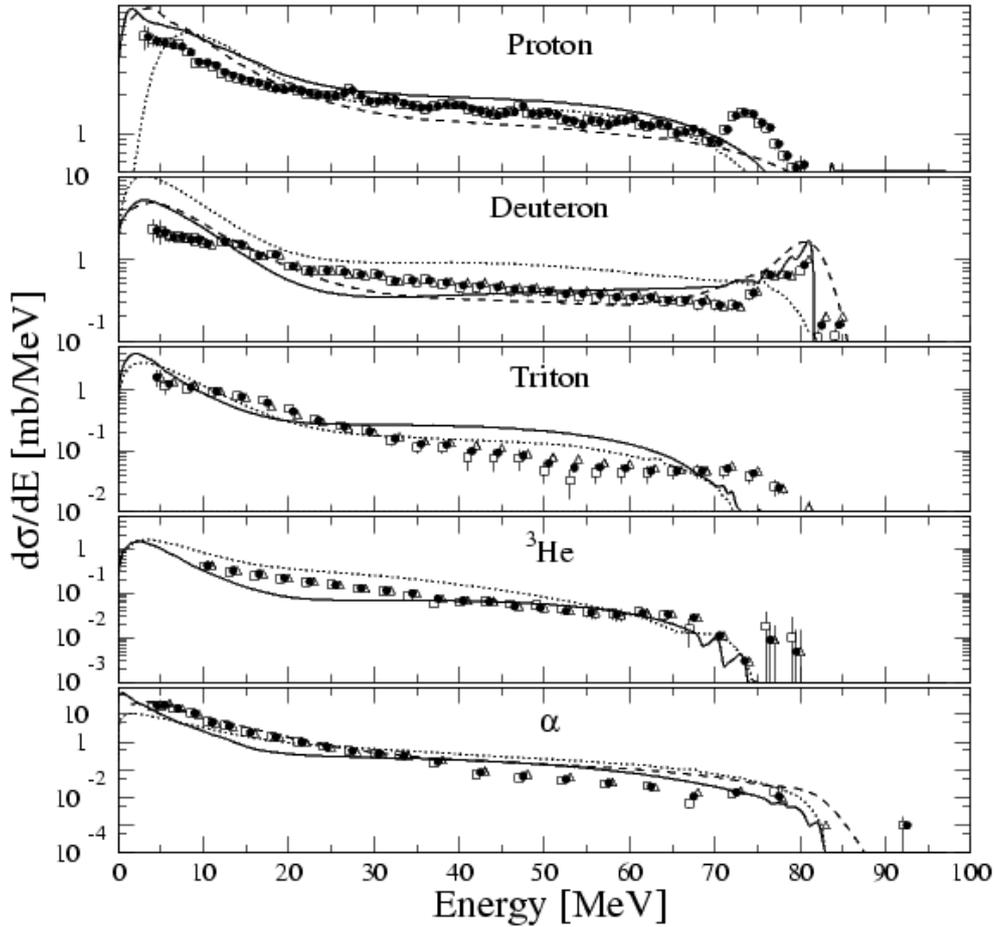

FIG. 17. Experimental energy-differential cross sections for neutron-induced $p$, $d$, $t$, $^3He$, and $\alpha$ production at 96 MeV. The open squares and open triangles represent the pure carbon and $CH_2$ data sets, respectively, whereas the filled circles represent the merging of the two sets. Note that the energy scales of the open triangles and open squares are shifted up and down by half an MeV for visibility. Curves indicate theoretical calculations based on GNASH I (dashed) and GNASH II (solid) and TALYS (dotted).

All calculations give a fair description of the energy dependence for all ejectiles. Both GNASH calculations overestimate the proton experimental data over the whole energy range, except the very highest energy, while TALYS calculations overestimate at intermediate energies. Moreover, the calculations for (n,p) reactions to discrete states underestimate the data. A study of the spectroscopic strengths for these states has done by Olsson *et al.* [51].

Concerning the deuteron spectra, all calculations give cross sections a factor of 2 or more larger than the experimental data in the evaporation peak. The TALYS code overestimates the data in the mid-energy range, 15-70 MeV, whereas both GNASH codes are in agreement with the spectrum part above 50 MeV, especially the high-energy deuteron peak.

The energy dependence of the triton spectrum is reasonably well described by the TALYS calculation, while GNASH II overestimates the data above about 30 MeV. Similar to the proton spectra, the calculations for the (n,t) reactions to discrete states underestimate the data.

For $^3$He particles, TALYS describes the spectral shapes over the entire energy range quite well, while the GNASH II curve falls below the data in the mid-energy range, 15-30 MeV.

In the case of α particles, TALYS and GNASH II tend to underpredict the low energy part, although GNASH I reproduces it well. All calculations overpredict the high-energy part of the spectrum.

Production cross sections are deduced by integration of the energy-differential spectra (see Table I), performed separately for the pure carbon, the $CH_2$, and the merged data sets. To be compared with the calculated cross sections, the experimental values in Table I have to be corrected for the undetected particles below the low-energy cutoff. Due to the high cutoff, this is particularly important for $^3$He. The cutoff correction is done using results from the model calculations which are given in Table I for the complete energy region and also below and above the cutoff.

In Table II we have given the cutoff-corrected production cross sections. Two methods have been used for the correction. In method I, the production cross section below the cutoff given by the model calculations below has simply been added to the experimental value. The given uncertainties remain unchanged in this case since we do not know the uncertainties of the model calculations. In method II we scale the experimental values by the ratio of the cross section below and above the cutoff as given by the model calculations in Table I. In this case, the experimental uncertainties are scaled accordingly.

TABLE I. Experimental production cross sections for protons, deuterons, tritons, $^3$He and α particles. Theoretical values resulting from GNASH I, GNASH II and TALYS calculations are given as well. The experimental data from the pure carbon, the $CH_2$ and the merged data sets are given in the second, third and fourth column, respectively. They have been obtained with cutoff energies of 3.0, 4.0, 5.0, 9.0 and 4.0 MeV for p, d, t, $^3$He and α particles, respectively. The fourth, fifth, and sixth column show results from the model calculations for the whole energy regions, and also below and above the experimental cutoff energies.

| $\sigma_{prod}$ | Experiment Production cross section (mb) | | | Model calculations Production cross section (mb) ( below/above the cutoffs) | | |
|---|---|---|---|---|---|---|
| | C | $CH_2$ | Merged | GNASH I | GNASH II | TALYS |
| (n,px) | 150 ± 15 | – | 149 ± 15 | 181.5 (23.6/157.9) | 208.5 (20.8/187.7) | 157.5 (1.7/155.8) |
| (n,dx) | 56 ± 6 | 57 ± 3 | 56 ± 3 | 85.3 (14.9/70.4) | 80.8 (15.3/65.5) | 145.2 (33.4/111.8) |
| (n,tx) | 19 ± 2 | 20 ± 1 | 20 ± 1 | – | 38.0 (14.4/23.6) | 33.5 (11.8/21.7) |
| (n,$^3$Hex) | 7.0 ± 1.0 | 7.7 ± 0.7 | 7.4 ± 0.6 | – | 13.2 (8.3/4.9) | 24.0 (8.5/15.5) |
| (n,αx) | 134 ± 14 | 135 ± 7 | 134 ± 8 | 248.2 (83.4/164.8) | 198.6 (139.1/59.5) | 103.0 (36.7/66.3) |

From the values in Table II we deduce a best estimate for the production cross sections which are given in Table III. For this best estimate, corrections using model calculations that seem too far off have been excluded while we have taken the average of the more reasonable ones. This procedure is somewhat ambiguous for the case of α-particle production. The uncertainties in this table are always the scaled experimental uncertainties of Table II.

TABLE II. Cutoff-corrected production cross sections from Table I. The corrections have been made using the GNASH I, GNASH II and TALYS calculations (see text).

| $\sigma_{prod}$ | Production cross section (mb) corrected for cutoff with method I | | | Production cross section (mb) corrected for cutoff with method II | | |
|---|---|---|---|---|---|---|
| | GNASH I | GNASH II | TALYS | GNASH I | GNASH II | TALYS |
| (n,px) | 173 ± 15 | 170 ± 15 | 151 ± 15 | 171 ± 17 | 166 ± 17 | 150 ± 15 |
| (n,dx) | 71 ± 3 | 71 ± 3 | 89 ± 3 | 68 ± 4 | 69± 4 | 72 ± 4 |
| (n,tx) | – | 34 ± 1 | 32 ± 1 | – | 32 ± 2 | 30 ± 2 |
| (n,$^3$Hex) | – | 15.7 ± 0.6 | 15.9 ± 0.6 | – | 19.9 ± 1.6 | 13.4 ± 1.1 |
| (n,αx) | 217 ± 8 | 273 ± 8 | 171 ± 8 | 202 ± 12 | 447 ± 27 | 203 ± 12 |

TABLE III. Best estimate of the cutoff-corrected production cross sections (mb) (see text).

| (n,px) | (n,dx) | (n,tx) | (n,$^3$Hex) | (n,αx) |
|---|---|---|---|---|
| 170 ± 17 | 70 ± 4 | 32 ± 2 | 16 ± 2 | 207 ± 12 |

The proton, deuteron, triton, $^3$He, and α-particle production cross sections, corrected by low-energy cutoffs which are given in Table III, are compared with the previous data at lower energies [2,53,54] in Fig. 18 (see figure caption for details). There seems to be a general agreement between the trends of the lower-energy data (open squares and stars) and the present data points (filled circles). However, there are notable differences from the previous publications [15,16,55] at the same neutron energy. The former are from the same data set, but with independent analysis work (open circles) [15,16] while the latter were from a different data set (filled triangles) [55]. The curves in this figure are based on GNASH I calculations [29].

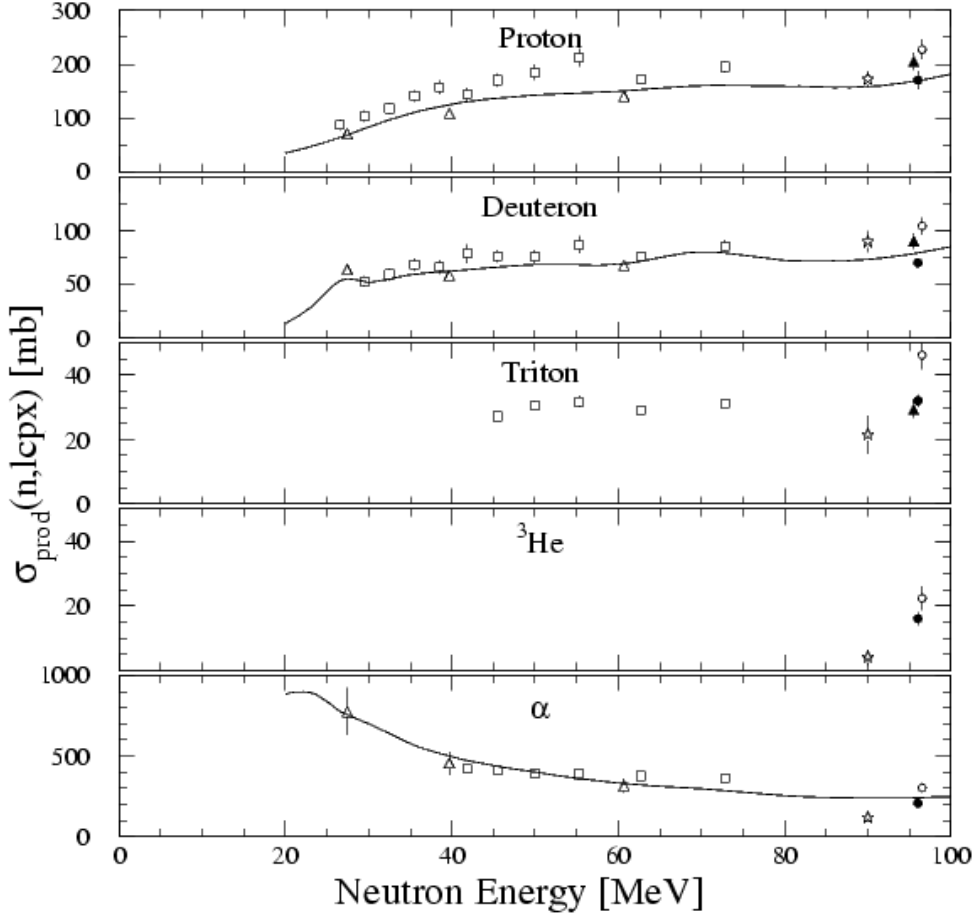

Fig. 18. Production cross sections, $\sigma_{prod}$, as a function of the incident neutron energy, $E_n$, for protons, deuterons, tritons, $^3$He and α particles from different measurements. The filled circles represent measurements from the present work as given in Table III, the filled triangles from a previous work [55] and the open circles from the same data set, but independent analysis work [15,16]. Note that the energy scales of the open circles and the filled triangles are shifted up and down by half an MeV for visibility. Open triangles are measurements from UC Davis [53], open squares from Louvain-la-Neuve [54], and open stars from Kellogg [2]. There seems to be general agreement between the trend of the previous data at lower energies and the present data point. The curves are based on GNASH I calculations [29].

## VII. Conclusions and outlook

In the present paper, we report an experimental data set for light-ion production induced by 96 MeV neutrons on carbon. Experimental double-differential cross sections ($d^2\sigma/d\Omega dE$) are measured at eight angles between 20° and 160°. Energy-differential ($d\sigma/dE$) and production cross sections are obtained for the five types of outgoing particles. To corroborate, we compare and then combine the data set with the carbon contribution extracted from a $(CH_2)_n$ target in a similar experiment on oxygen and silicon [12,17]. The combined spectra have improved the statistical accuracy and displayed consistency in shape and magnitude. The double-differential and energy-differential cross sections from different target data sets are in very good agreement both in magnitude and shape. The production cross sections differ within 5-10%, mainly from the systematic uncertainties. However, we find that there are 30-40% differences from a previous publication [15] in which other analysis procedures were used and in which corroboration of the type presented here was missing.

In general, theoretical calculations based on nuclear reaction codes including direct, pre-equilibrium and statistical calculations predict a fair account of the magnitude of the experimental cross sections. For proton emission, the shape of the spectra for the double-differential and energy-differential cross sections are reasonably well described. However, there are significant differences between theory and experiment concerning the magnitude and the shape of the spectra for the complex ejectiles. This may not be so surprising, since two well-known unsolved aspects of nuclear reaction theory meet here: (a) the application of statistically based models such as the optical model, the Hauser-Feshbach model and level densities on a system of only 12 nucleons, and (b) a sound theoretical description of complex-particle pre-equilibrium emission.

Using MEDLEY at the new Uppsala neutron beam facility [56], we plan to measure double-differential cross sections for light-ion production on oxygen, silicon, iron, lead, bismuth and uranium at 175 MeV. Thus far, we have collected data on $^{12}$C(n,lcp) and presented preliminary double-differential cross sections at the ND2007 conference [57].

**Acknowledgements**

This work was supported by the Swedish Natural Science Research Council, the Swedish Nuclear Fuel and Waste Management Company, the Swedish Nuclear Power Inspectorate, Ringhals AB, and the Swedish Defence Research Agency. The authors wish to thank the The Svedberg Laboratory for excellent support. Y.W. is grateful to the scientific exchange program between the Japan Society for the Promotion of Science and the Royal Swedish Academy of Sciences.